\documentclass[aps,twocolumn]{aastex61}

\usepackage{url}
\usepackage{enumitem}

\newcommand{\hMsun}{{\ifmmode{h^{-1}{\rm
        {M_{\odot}}}}\else{$h^{-1}{\rm{M_{\odot}}}$~}\fi}}
\newcommand{\hMpc}{{\ifmmode{h^{-1}{\rm Mpc}}\else{$h^{-1}$Mpc }\fi}}
\def\be{\begin{equation}}
\def\ee{\end{equation}}
\def\ba{\begin{eqnarray}}
\def\ea{\end{eqnarray}}

\received{November 6, 2017}
\revised{November 27, 2017}
\accepted{\today}
\submitjournal{ApJ}

\shorttitle{Cosmology from LSS deep learning}

\begin{document}

\title{Cosmological parameter estimation from large-scale structure deep learning}

\author{Shuyang Pan, Miaoxin Liu}
\affiliation{School of Physics and Astronomy, Sun Yat-Sen University, Guangzhou 510297, P.R.China}

\author{Jaime Forero-Romero}
\affiliation{Departamento de F{\'i}sica, Universidad de los Andes, Cra. 1 No. 18A-10 Edificio Ip, CP 111711, Bogot{\'a}, Colombia}

\author{Cristiano G. Sabiu}
\affiliation{Department of Astronomy, Yonsei University, Seoul, Korea}

\author{Zhigang Li}
\affiliation{College of Physics and Electronic Engineering, Nanyang Normal University, Nanyang, Henan, 473061, China}

\author{Haitao Miao}
\affiliation{School of Physics and Astronomy, Sun Yat-Sen University, Guangzhou 510297, P.R.China}
\author{Xiao-Dong Li$\ ^\ast$}
\affiliation{School of Physics and Astronomy, Sun Yat-Sen University, Guangzhou 510297, P.R.China}

\email{$\ ^\ast$ Corresponding Author: lixiaod25@mail.sysu.edu.cn}




\begin{abstract}
We propose a light-weight deep convolutional neural network (CNN) to estimate the
cosmological parameters from simulated 3-dimensional dark matter distributions with high accuracy.
The training set is based on 465 realizations of a cubic box with a side length of $256\ h^{-1}\ \rm Mpc$, 
sampled with $128^3$ particles interpolated over a cubic grid of $128^3$ voxels.
These volumes have cosmological parameters varying within the flat $\Lambda$CDM parameter space of 
$0.16 \leq \Omega_m \leq 0.46$ and $2.0 \leq 10^9 A_s \leq 2.3$.
The neural network takes as an input cubes with $32^3$ voxels and has three convolution layers, 
three dense layers, together with some batch normalization and pooling layers. 
In the final predictions from the network we find a $2.5\%$ bias on the primordial amplitude $\sigma_8$ 
that can not easily be resolved by continued training.
We correct this bias to obtain unprecedented accuracy in the cosmological parameter estimation 
with statistical uncertainties of $\delta \Omega_m$=0.0015 and $\delta \sigma_8$=0.0029,
which are several times better than the results of previous CNN works.
Compared with a 2-point analysis method using clustering region of
0-130 and 10-130 $h^{-1}$ Mpc, 
the CNN constraints are several times and an order of magnitude
more precise, respectively.
Finally, we conduct preliminary checks of 
the error-tolerance abilities of the neural network,
and find that it exhibits robustness against smoothing, masking, random noise, 
global variation, rotation, reflection, and simulation resolution.
Those effects are well understood in typical clustering analysis, 
but had not been tested before for the CNN approach. 
Our work shows that CNN can be more promising than people expected 
in deriving tight cosmological constraints from the cosmic large scale structure.
\end{abstract}



\section{Introduction}

The current standard model of cosmology has been highly successful at describing
the Universe on large scales. 
From the anisotropic temperature fluctuations in the cosmic microwave background 
(CMB) to the late time clustering of galaxies,
the vacuum energy dominated cold dark matter model ($\Lambda$CDM)
\citep{weinberg1989cosmological,peebles2003cosmological,miao2011dark}
fits the data surprisingly well \citep{riess1998observational,perlmutter1999measurements,weinberg2013observational,
ade2016planck,alam2017clustering}.
For cosmologists, one main task would be to precisely estimate the parameters of
the Universe, such as the dark matter ratio $\Omega_m$, the local expansion rate 
$H_0$, the amplitude and index of the primordial fluctuation $A_s$ and $n_s$,
the dark energy equation of state $w$ together with its time dependence $w_a$, and
so on.

The spatial distribution of galaxies on scales of a few
hundred Megaparsecs (Mpc) forms a distinct, very complicated filamentary motif  known as the `cosmic web'
\citep{1986ApJ...304...15B,1986ApJ...302L...1D,
  2012ApJS..199...26H,2004ApJ...606..702T,2014A&A...566A.108G}.
The distribution and clustering properties of galaxies in the cosmic web
encodes information on the expansion and the structure growth history of the Universe. 
In the next decades, several large scale surveys
(e.g., DESI\footnote{https://desi.lbl.gov/}, EUCLID\footnote{http://sci.esa.int/euclid/}, LSST\footnote{https://www.lsst.org/},
WFIRST\footnote{https://wfirst.gsfc.nasa.gov/})
will begin operations to map out an unprecedented large volume of the Universe with extraordinary precision. 
It becomes essential to develop powerful tools that can
comprehensively and reliably infer the cosmological parameters
from large scale structure (LSS) data.


Currently, the most widely-adopted LSS data mining methods is still the
2-point correlation function (2pCF) or power spectrum measurements,
which are sensitive to the geometric and structure growth history of the Universe
\citep{kaiser1987clustering,ballinger1996measuring,Eisenstein:1998tu,2003ApJ...594..665B,2003ApJ...598..720S}.
These methods have achieved tremendous success when applied to
a series of galaxy redshift surveys such as the 2-degree Field Galaxy Redshift Survey
(2dFGRS; \cite{2df:Colless:2003wz}),
the 6-degree Field Galaxy Survey (6dFGS; \cite{beutler20116df}),
the WiggleZ survey \cite{blake2011wigglez,blake2011wigglezb},
and the Sloan Digital Sky Survey (SDSS;
\cite{york2000sloan,Eisenstein:2005su,Percival:2007yw,
anderson2012clustering,sanchez2012clustering,sanchez2013clustering,
anderson2014clustering,samushia2014clustering,ross2015clustering,
beutler2016clustering,sanchez2016clustering,
alam2017clustering,chuang2017clustering}.
The main caveat of this method is that,
the distribution of structures and their velocities on scales of $\lesssim40h^{-1}\ \rm Mpc$ are highly affected by the non-linear processes, making it difficult to conduct a comparison between observations and theories.

Ongoing research seeks to utilise LSS data on non-linear scales or beyond the usual 2nd order spatial statistics. 
The next order correlation function, the 3-point correlation function, has been shown to add cosmological constraints beyond the 2pCF \citep{2017MNRAS.469.1738S} and it has also shown promise in constraining modified gravity models \citep{2016A&A...592A..38S}. 
The 4-point function may also lead to improved constraints if it can be modelled correctly \citep{2019ApJS..242...29S}.

Some other tests include the proposal to use the apparent stretching of cosmic
voids as a probe of geometry \citep{ryden1995measuring,lavaux2012precision}; 
the redshift invariance of the comoving scale information in the LSS to
probe the expansion history \citep{Li2014,Li2017}; the symmetry
properties of galaxy pairs to conduct an Alcock-Paczynski (AP) tests \citep{ap,marinoni2010geometric}; the redshift-dependent property of the
AP effect to overcome the effect of redshift space distortion (RSD) \citep{Li2014,Li2015} to successfully derive tight dark energy constraints from the SDSS galaxies \citep{Li2016,Li2018,Li2019,Zhang2019}.
Recently, \cite{Fang2018} applied the so-called $\beta$-skeleton statistics to study LSS and proposed its application for cosmological analysis;
\cite{KR2018} proposed to use the large-scale Bayesian inference framework
to constrain parameters via the AP test.

To summarize, there are many alternative ideas and concepts that have been used proposed and used to extract information from the LSS, 
and one may refer to \cite{weinberg2013observational} and the references therein for a more complete overview.

While cosmologists have obtained prominent information about the physics of the
Universe via the current statistical methods, due to the extreme sophistication of 
the cosmic web we are still far  from having a statistical method to 
comprehensively explore the overwhelming information encoded in the cosmic LSS.
Fortunately, recent  developments in machine learning techniques may allow us to
capture and extract more cosmological information from the complex LSS data.

Machine learning techniques, especially the deep learning algorithms 
based on deep neural networks,
are becoming a mainstream toolkit for modeling the relationship
between complex data and the underlying variables that it corresponds to.
They make it possible to extract and analyze features contained within the data,
which can not be easily identified via traditional  methods of scientific research
\footnote{\url{https://www.oreilly.com/ideas/a-look-at-deep-learning-for-science}}.
Recently, machine learning techniques have been applied to many sub-fields of cosmology, 
including weak gravitational lensing \citep{Schmelzle:2017vwd,
Gupta:2018eev,Springer:2018aak,Fluri:2019qtp,Jeffrey:2019fag,Merten:2018bgr,Peel:2018aei,Tewes:2018she}, 
the cosmic microwave background \citep{Caldeira:2018ojb,Rodriguez:2018mjb,Perraudin:2018rbt,Munchmeyer:2019kng,Mishra:2019sep},
the large scale structure \citep{Soumagnac:2013odh,Lucie-Smith:2018smo,
Modi:2018cfi,Berger:2018aey,He:2018ggn,Lucie-Smith:2019hdl,
Pfeffer:2019pca,Ramanah:2019cbm,Troster:2019mys,Zhang:2019ryt}, 
gravitational waves \citep{Dreissigacker:2019edy,Gebhard:2019ldz}, 
cosmic reionization \citep{LaPlante:2018pst,Gillet:2018fgb,Hassan:2018bbm,Chardin:2019euc,Hassan:2019cal},
supernovae \citep{Lochner:2016hbn,Moss:2018tug,Ishida:2018uqu,Li:2019ybe,Muthukrishna:2019wpf}.
For more details, one can refer to \cite{Mehta:2018dln,Jennings:2018eko,Carleo:2019ptp,Ntampaka:2019udw} and the references therein.

In a pioneering work, \cite{Ravanbakhsh:2017bbi} presented a CNN (convolutional neural network) to infer cosmological parameters from simulated 3-dimensional dark matter density fields.
They were able to constraint $\Omega_m$ and the matter over-density variance 
$\sigma_8$, finding that the machine learning techniques can outperform the
traditional 2pCF statistics.
\cite{Mathuriya:2018luj} presented a more sophisticated framework, 
which can achieve synchronous parallel calculation on tens of thousands of nodes,
and simultaneously predict $\Omega_m$, $\sigma_8$, and the primordial power spectrum index $n_s$.

In this work we build upon those previous studies to explore 
a new deep learning architecture
and perform new tests to study the LSS. 
We show that it is possible to constrain $\Omega_m$ and $\sigma_8$ using 
$32^3$ voxels only as an input, a small number compared to the larger sizes of $64^3$ and $128^3$ used by \cite{Ravanbakhsh:2017bbi} and \cite{Mathuriya:2018luj}, respectively. 
Compared with \cite{Ravanbakhsh:2017bbi}, 
we achieve an order of magnitude better constraints on the parameters,
while the architecture we propose is also simpler than the ones suggested in those two
works.
Finally, although CNNs are able to achieving state-of-the-art performance on many tasks, 
some recent  studies revealed that they can also be easily fooled \cite{Moosavi} by either giving wrong prediction
from minor changes in the inputs or giving seemingly correct values for unreasonable inputs.
Here we also test for error-tolerance abilities of the neural network to different effects 
that are well understood in traditional clustering analysis 
(smoothing, masking, random noise, global variation, rotation, reflection, simulation resolution) 
but have not been fully explored in the context of predicting cosmological parameters using CNNs.

This paper is structured as follows.
In Section 2 we introduce the samples used for the training and testing,
while the Section 3 we explain the architecture of our neural network.
The results are presented in Section 4.
We conclude in Section 5 by discussing the future of the technique and its caveats.

\section{Data}

The training and testing samples are created with the COmoving Lagrangian
Acceleration (COLA) code  \citep{Tassev:cola,Koda:2015mca},
which is designed as a mixture of N-body and perturbation theory 
to  simulations with fast speed and good accuracy,
We choose COLA because it is hundreds of times faster than N-body simulations,
while keeping a good accuracy in generating structures on non-linear scales.

We change two cosmological parameters in our simulations, the fraction of matter, $\Omega_m$,
and the amplitude of the primordial power spectrum, $A_s$.
Values of the other parameters are taken as $\Omega_b=0.048206$, $h=0.6777$, 
$n_s=0.96$, the same as the MultiDark Planck N-body simulations 
\citep{klypin2016multidark}.

We vary the values of $\Omega_m$ and $A_s$ on a $31\times15$ grid, i.e.
$0.16 \leq \Omega_m \leq 0.46$ with step size 0.01,
and $2.0 \leq 10^9 A_s \leq 2.3$ with step size 0.02.
This parameter space is centered around the Planck 2015 best fit cosmology \citep{ade2016planck}
\footnote{Planck 2015 (TT,TE,EE+lowP+lensing) gives
$\Omega_m = 0.3121 \pm 0.0087$,
$10^9 A_s = 2.13 \pm 0.053$,
$\sigma_8 = 0.8150 \pm 0.0087$ in the $\Lambda$CDM framework.}.
This leads to a varying $\sigma_8$ in the range of 0.4-1.1.

For all samples, we run a simulation with $128^3$ particles,
in a $(256$ $h^{-1} \rm Mpc)^3$ box, using 40 timesteps.
We output the normalized density field,
\begin{equation}
 \delta \rho({\bf x}) \equiv \frac{\rho({\bf x})}{\bar\rho},
\end{equation}
on a grid with $128^3$ voxels at redshift $z=0$, 

To train the neural network we generate $31\times15$ samples (i.e. boxes) -- one sample
for an individual cosmology.
The simulation adopted the second-order Lagrangian perturbation theory
(2LPT) initial conditions at $z_i=39$.
Each cosmology is evolved from initial conditions with different random seeds
and thus different distributions of large scale power,
so that our neural network can capture the cosmic variance.

To test the neural network we generated two sets of testing samples:
\begin{itemize}
 \item The ``single-cosmology'' testing samples,
for which we generated 500 samples sharing the same cosmology $(\Omega_m, \sigma_8)=(0.3072,0.8228)$.
This allows us to validate the statistical error of the neural network predictions.
 \item The ``multi-cosmology'' samples,
wherein we have $31\times15$ samples, using different cosmologies (on the same grid of the testing sample cosmology grid).
The multi-cosmology set allows us to validate the accuracy of the parameter estimation in the whole parameter space.
\end{itemize}
The testing samples are created using initial conditions different from those of the training samples.

\begin{figure*}
   \centering{
   \includegraphics[width=12cm]{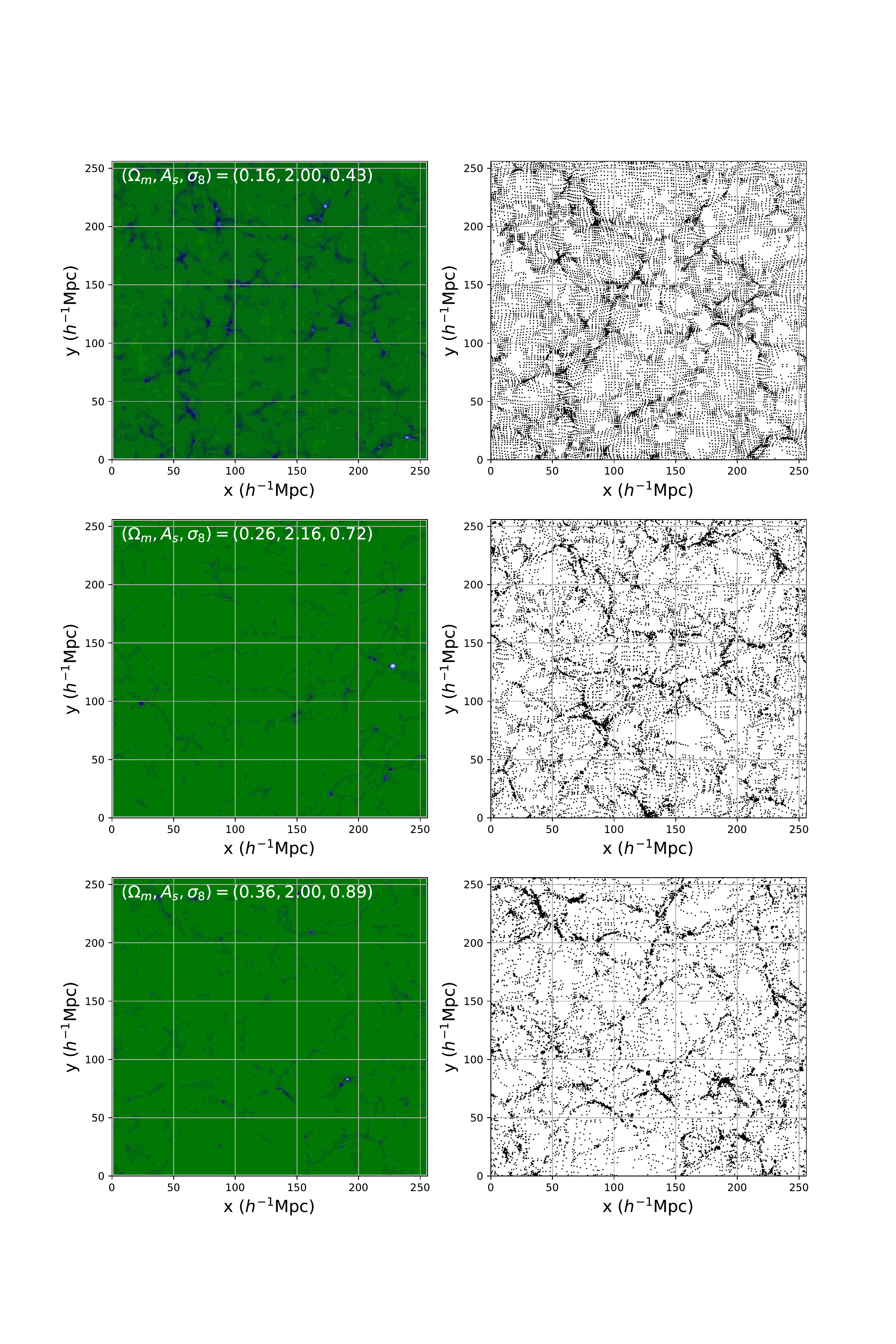}
   }
   \caption{\label{fig_field}
   The density field (left) and particle distribution (right) in three cosmologies
    $(\Omega_m, A_s, \sigma_8)=(0.16,2,0.43),\ (0.26, 2.16, 0.72), (0.36, 2.0, 0.89)$,
    selected from the training sample.
   We plot the 2D distribution, with the third dimension restricted to a thin slice $0 h^{-1} {\rm Mpc}<z<2 h^{-1} {\rm Mpc}$.
   The clustering strength is enhanced when increasing $\Omega_m$ or $A_s$,
   making the structures more ``compact''.
   We train neural networks to build up connections between the density fields
   and their underlying cosmological parameters.
   }
\end{figure*}
\begin{figure}
   \centering
    \includegraphics[height=8cm]{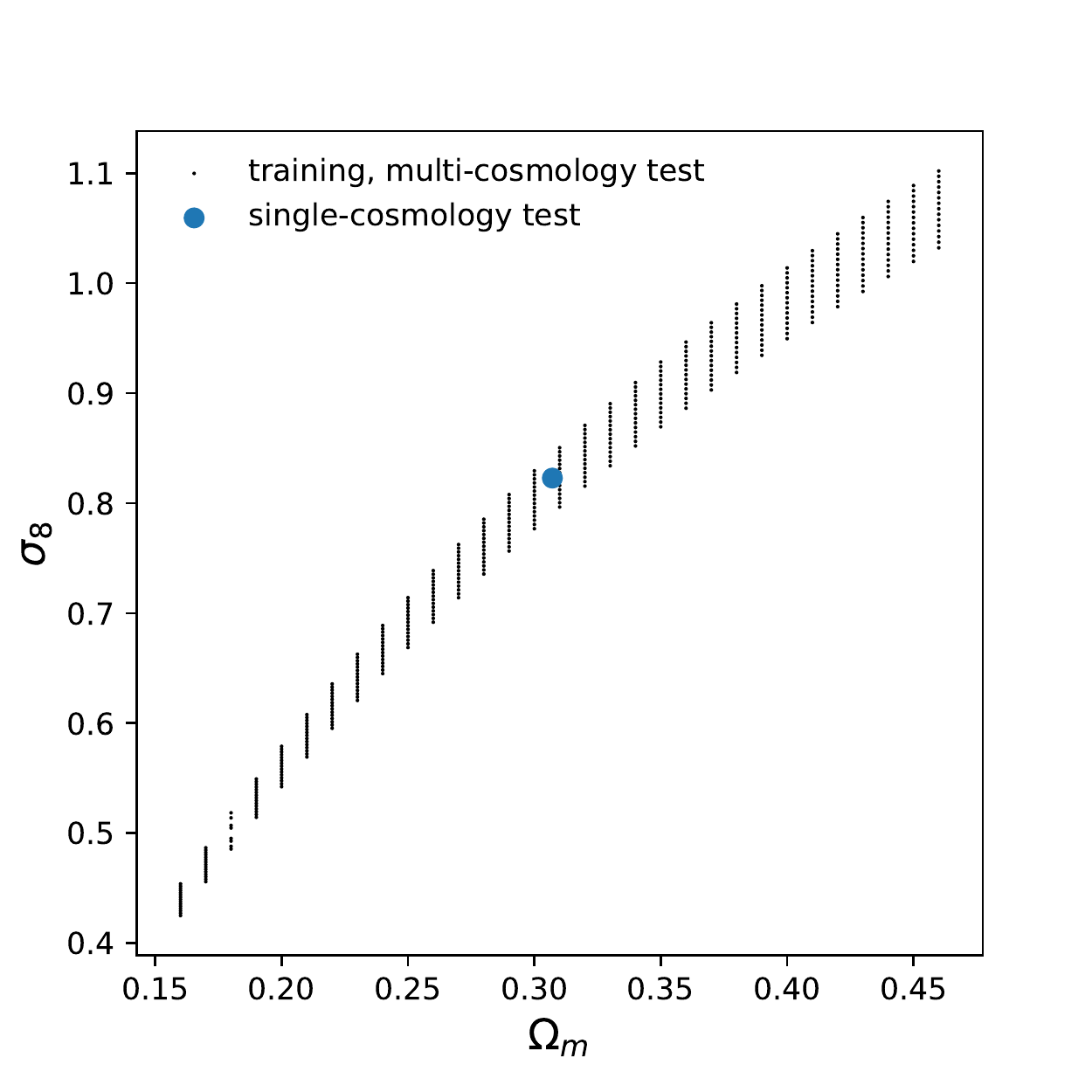}
   \caption{\label{fig_train_grid}
   $\Omega_m$ and $\sigma_8$ values for the 465 training samples and single-cosmology test
   samples.
   The multi-cosmology test samples have exactly the same values of $\Omega_m$ and $\sigma_8$ as those in training samples.
   }
\end{figure}

In Figure \ref{fig_field} we plot the density fields and the particle distributions
of three training samples,
$(\Omega_m, A_s, \sigma_8)=(0.16,2,0.43),\ (0.26, 2.16, 0.72), (0.36, 2.0, 0.89)$.
Obviously, the clustering strength increases when increasing $\Omega_m$ or $A_s$,
making the structures more compact.
In Figure \ref{fig_train_grid} we plot the cosmologies of the
training and testing samples, in the $\Omega_m$-$\sigma_8$ space.
In contrast to the $\Omega_m$-$A_s$ space, here we see a strong degeneracy between the two
parameters. The prior distributions of these parameters may influence the performance of the CNN
training and predicting, and this influence is unchecked in this work.
The prior adopted in this work is uniformly distributed in $\Omega_m$-$A_s$ 
space which exhibits  a strong degeneracy in $\Omega_m$-$\sigma_8$, which may not be optimal.

\begin{figure*}
   \centering{
    \includegraphics[width=16cm]{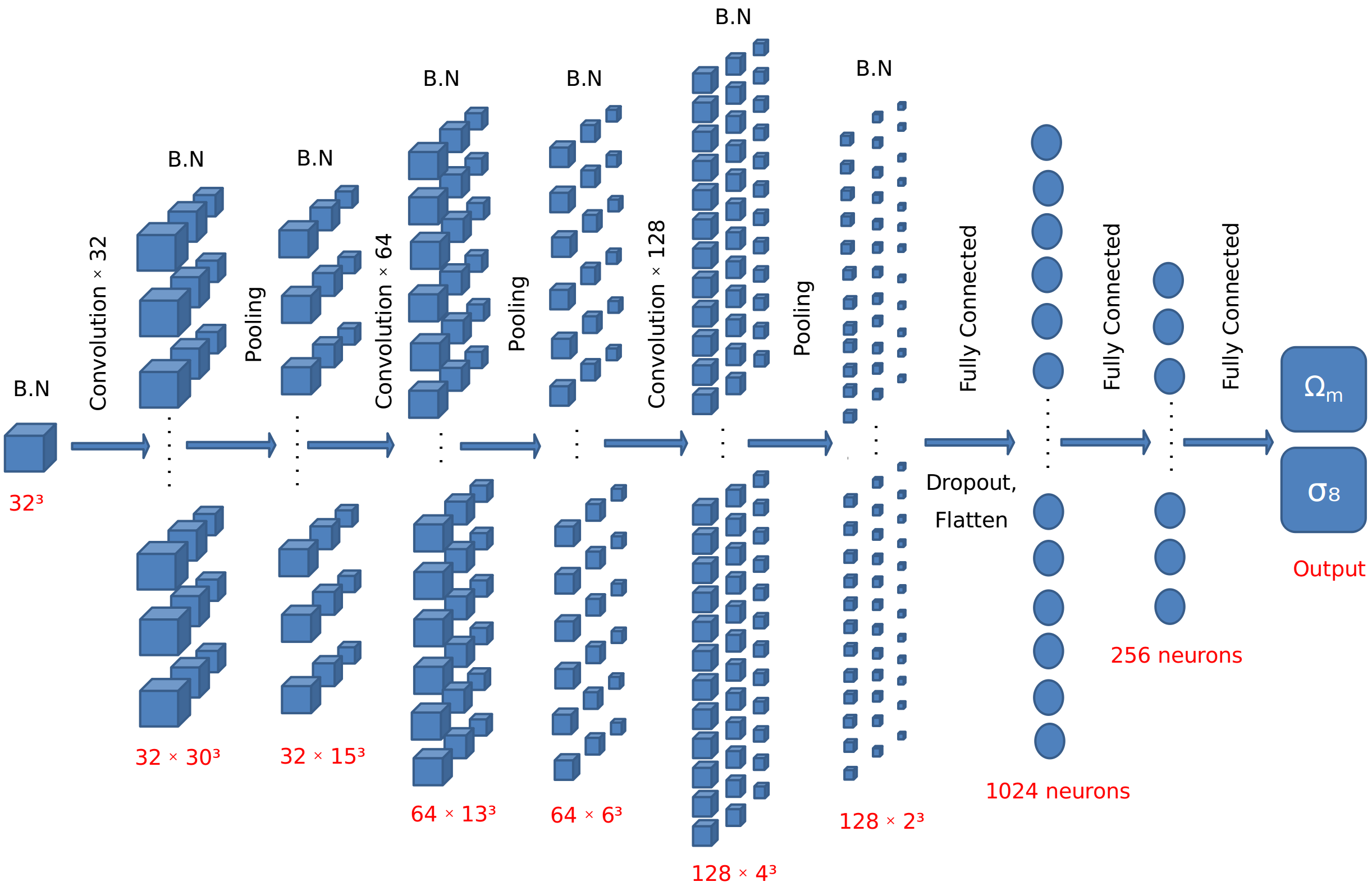}
   }
   \caption{\label{fig_archi}
   The architecture of our neural network.
   A cube having $32^3$ voxels is fed to the network.
   The three convolution layers have 32, 64, 128 filters, respectively.
   Beside each convolution layer, a batch normalization layer is added before it to
   normalize the distribution (so that to enhance the stability),
   and a pooling layer is placed after it to decrease the size of the output.
   After that, we got $128\times 2^3$ voxels containing the extracted features.
   They are then converted to a 1-d vector by the flatten layer,
   and passed to three dense layers with 1028, 24, 2 neurons,
   to output the final predictions of $\Omega_m$ and $\sigma_8$.
   }
\end{figure*}

\begin{figure*}
   \centering
    \includegraphics[width=15cm]{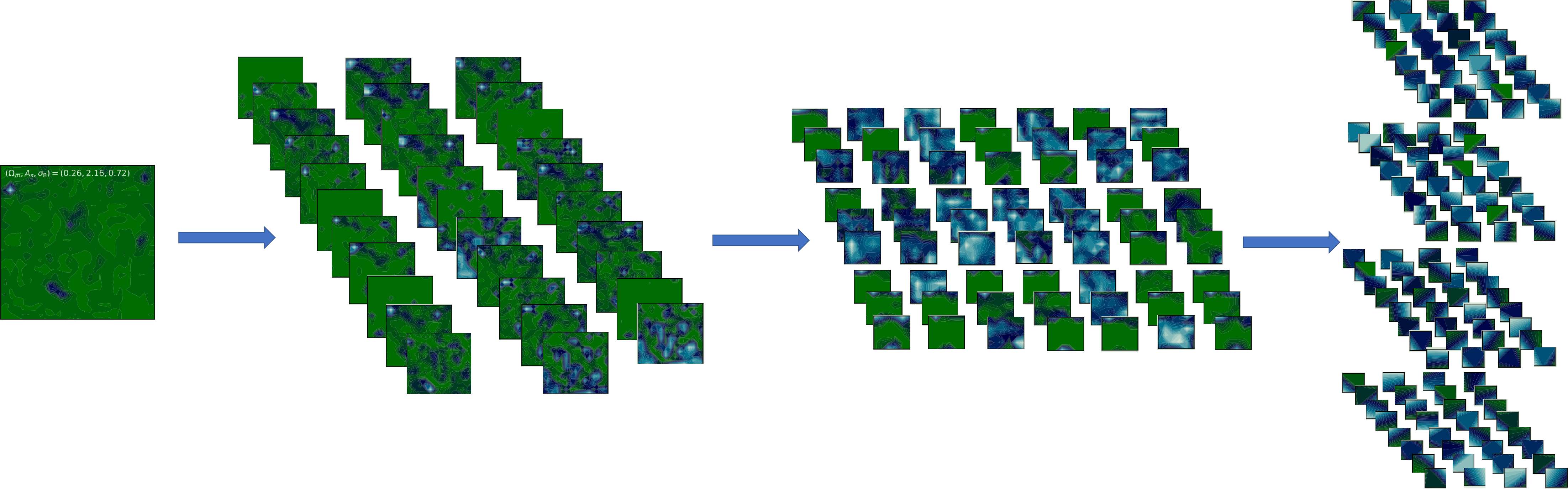}
   \caption{\label{fig_layers1}
   Layer-by-layer outputs of the CNN when fed by
   a sample with cosmology parameters $(\Omega_m, A_s, \sigma_8)=(0.26, 2.16, 0.72)$.
   The many filters,
   determined by the 896/55,360/221,312 trainable parameters in the three convolutions layers,
   can capture various types of features.
   The final outputs of the CNN is a set of 128 $2^3$-boxes containing the most compressed features extracted from the data.
   They are passed to the dense layers (not plotted here) for parameter estimation.
   }
\end{figure*}

\begin{figure*}
   \centering
    \includegraphics[width=15cm]{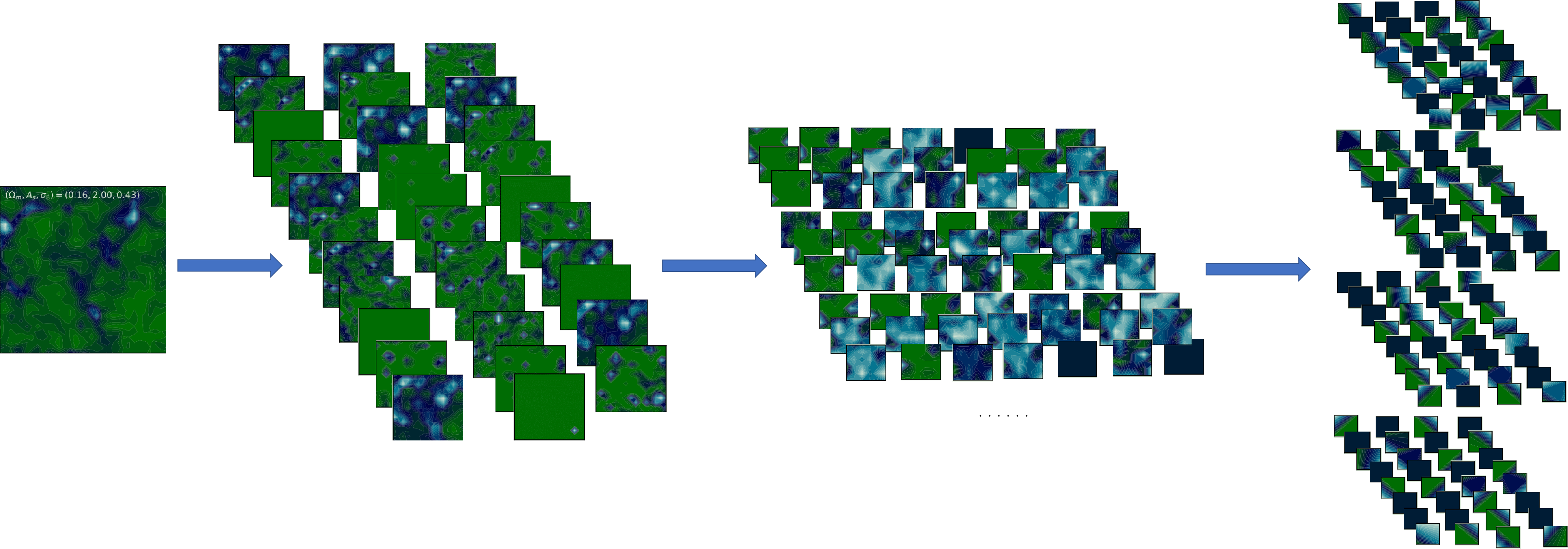}
   \caption{\label{fig_layers2}
   Same as Figure \ref{fig_layers1}, except that for the case of $(\Omega_m, A_s, \sigma_8)=(0.26, 2.00, 0.43)$.
   The features extracted are significantly different from those in the cosmology $(\Omega_m, A_s, \sigma_8)=(0.26, 2.16, 0.72)$,
   making it possible to distinguish these two cosmologies.
   }
\end{figure*}

\begin{figure*}
   \centering
    \includegraphics[width=16cm]{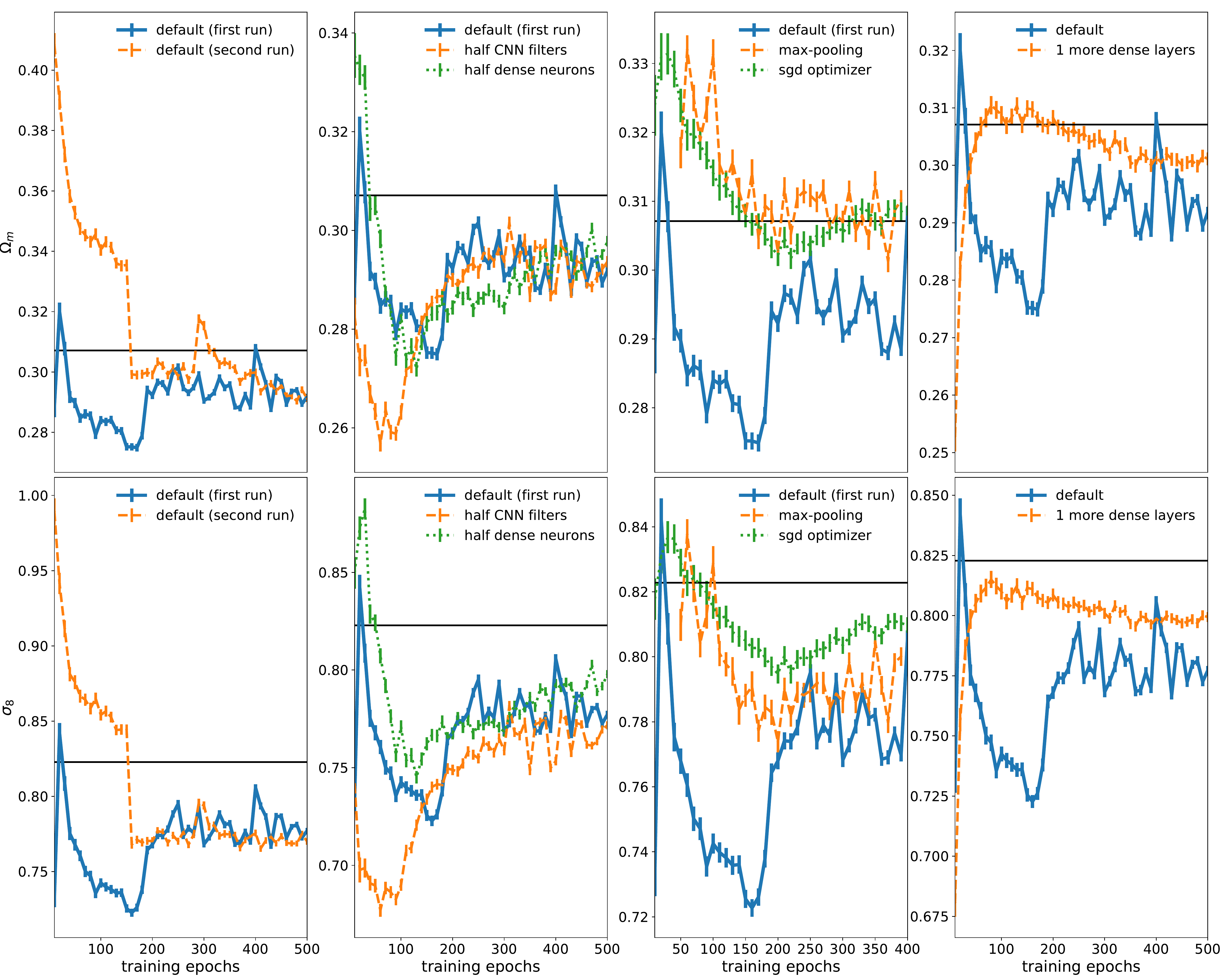}
   \caption{\label{fig_lc}
   Learning curve using different architectures.
   First panel: two runs using the default options reaches convergence after 160 epochs.
   Second panel: decreasing the number of CNN filters or dense neurons by 50\%,
   no significant change in the performance.
   Third panel: among our trials of different options,
   using max-pooling or sgd optimizer can notably enhances the performance.
   Fourth panel: an extra dense layer with 512 neurons are added before the final outputs
   to achieve a more accurate mapping from CNN outputs to the cosmological parameters.
   A good performance is detected at $\approx$80 epochs; more training epochs results in over-fitting.}
\end{figure*}

\begin{figure*}
   \centering
    \includegraphics[width=16cm]{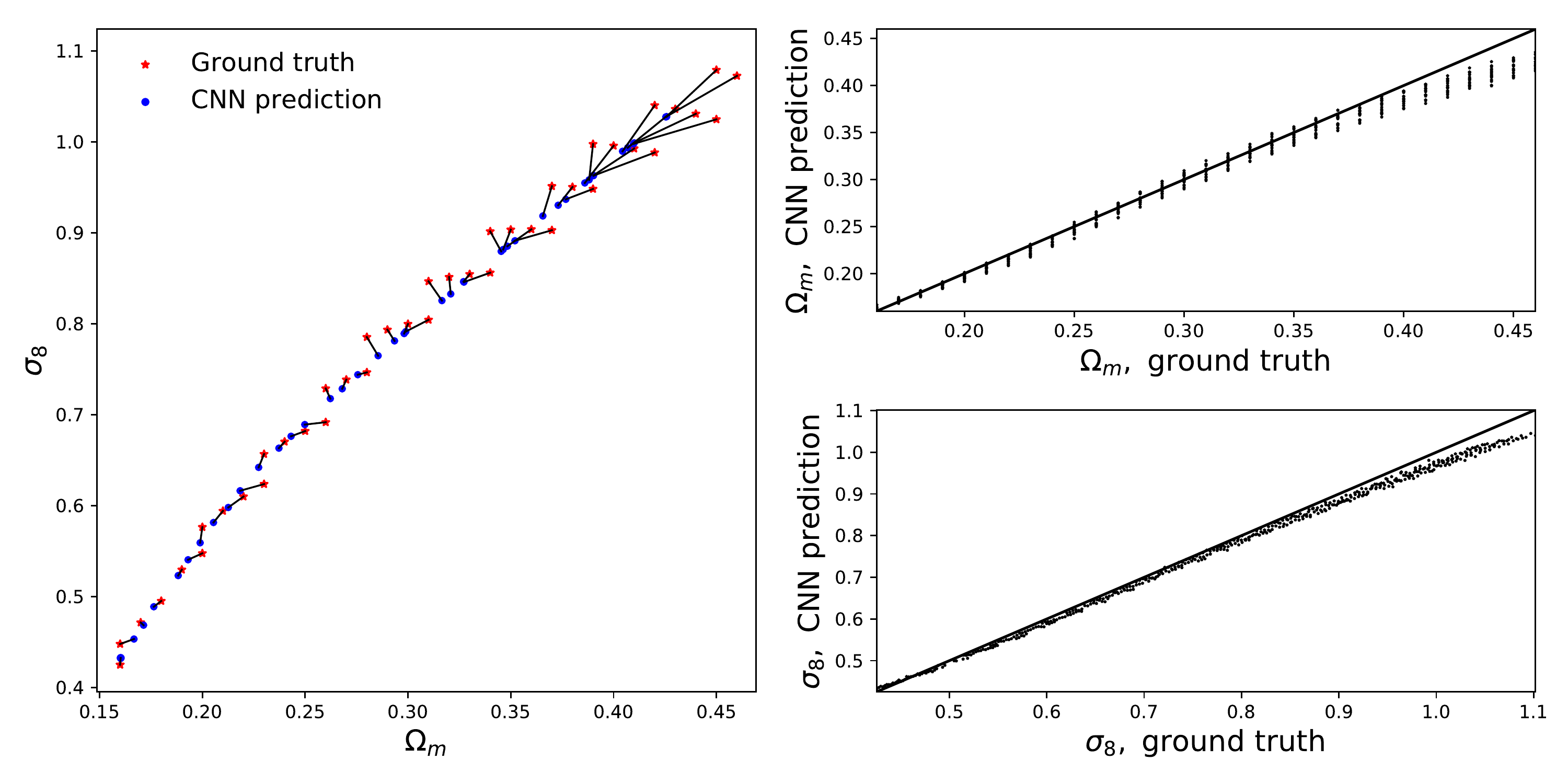}
   \caption{\label{fig_multi_cosmology}
   Test of a CNN architecture (sgd) on a multi-cosmology grid.
   There is a strong degeneracy between $\Omega_m$ and $\sigma_8$.
   {\it Left panel}: Ground truth and CNN predictions of $\Omega_m$ and $\sigma_8$, in the 2-d parameter space.
   The black lines show the difference between them.
   The bias is larger at the upper-right corner of the parameter space.
   {\it Right panels}: Ground truth and CNN predictions for $\Omega_m$ and $\sigma_8$ panels, respectively.
   }
\end{figure*}

\begin{figure*}
   \centering
    \includegraphics[width=16cm]{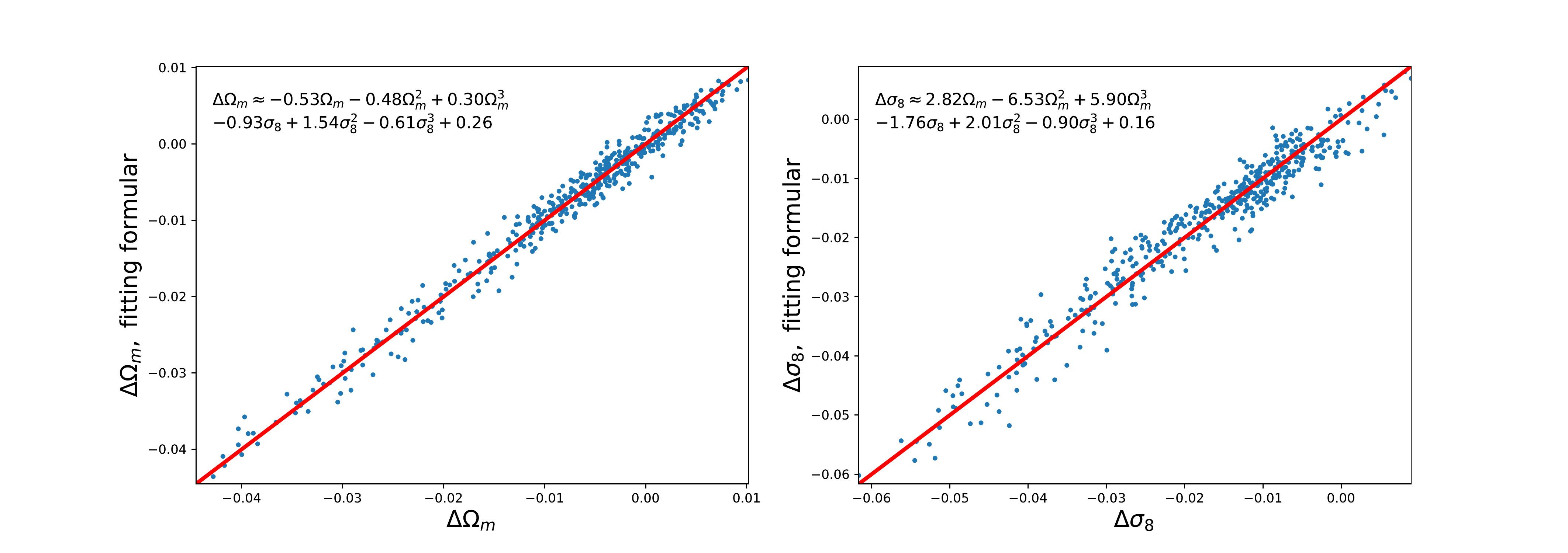}
   \caption{\label{fig_bias}
   Distribution of the systematic bias in the CNN predicted $\Omega_m$ and $\sigma_8$ (denoted as $\Delta \Omega_m$ and $\Delta \sigma_8$).
   Very roughly, in the parameter space we studied,
    there is $|\Delta \Omega_m| \lesssim 0.03$ and $|\Delta \sigma_8| \lesssim 0.05$,
    with mean value of $\bar|\Delta \Omega_m|=0.01$ and $\bar|\Delta \sigma_8|=0.018$.
   In practice one can calibrate the results by subtract the systematic bias in the CNN predictions (e.g., using the fitting formula shown in the panels),
   making the final estimation unbiased.
   }
\end{figure*}

\begin{figure*}
   \centering
    \includegraphics[width=16cm]{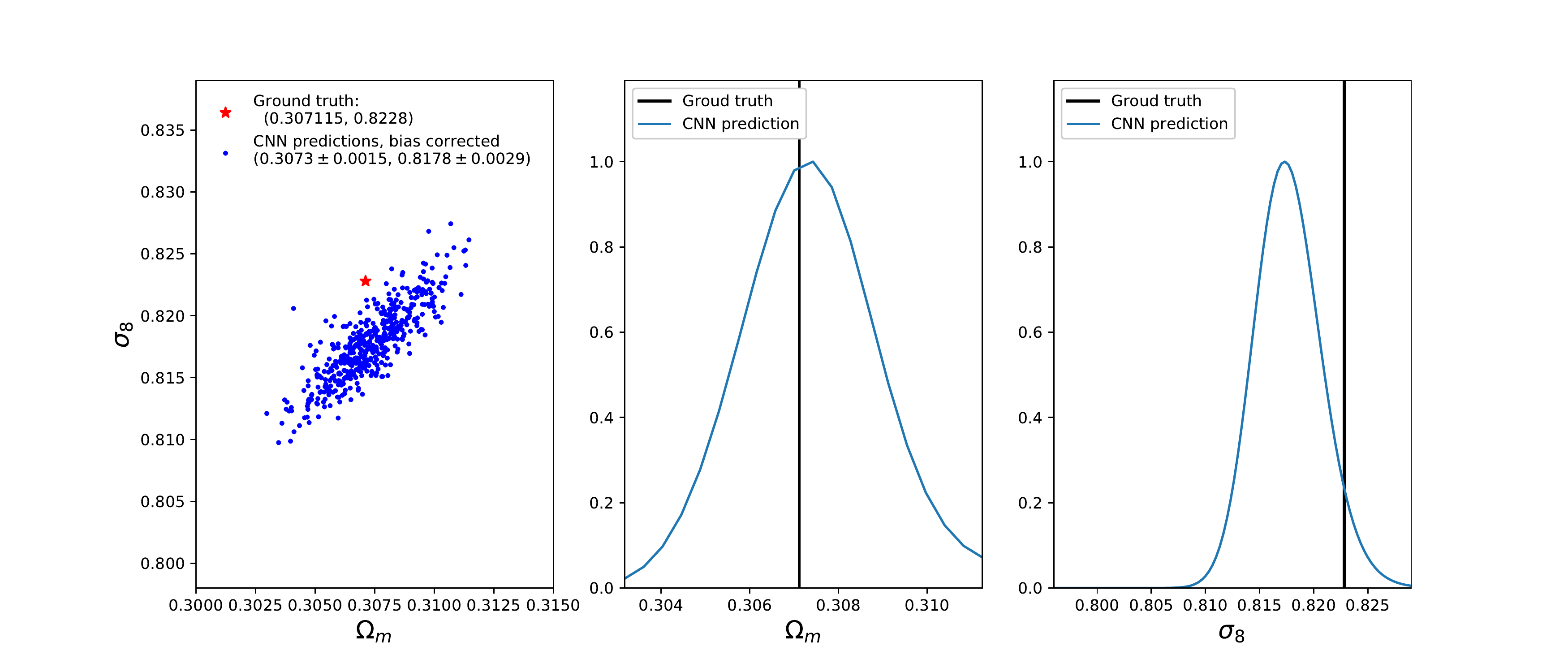}
   \caption{\label{single_cosmology}
   Test of a CNN architecture (sgd) on the single-cosmology samples.
   {\it Left panel}: Ground truth (red star) and CNN predictions (blue dots) of $\Omega_m$ and $\sigma_8$, in the 2-d parameter space.
   The CNN well predicts the values of $\Omega_m$, but has a bias in estimating $\sigma_8$.
   {\it Middle and Right panels}: Likelihood distribution of $\Omega_m$, $\sigma_8$ from the CNN predictions.
   }
\end{figure*}

\section{Methodology}
\label{sec:method}

One disadvantage of deep learning is that it is almost impossible to design an architecture
from first principles for the task at hand.
Furthermore, although  a precise parameter estimation is achieved,
it is difficult to say what spatial scale or features are the most relevant
to predict the final cosmological parameters.
Here we use a large number of filters for the initial spatial convolution,
based on the belief that small scale structures contain abundant information
and should be convolved by many filters to extract various features.

The input of the whole network is a $32^3$-voxel
(i.e. $(64 {h^{-1}\rm\ Mpc})^3$) subcube of the original density fields that is stored in
a $128^3$-voxel cube.
We do not feed the whole $128^3$ voxel cube to the neural network based on three considerations.
\begin{enumerate}
\item To learn a larger cube the network should have more neurons or layers
and thus its training becomes much more difficult and expensive.
\item Large cubes is challenging for the memory especially for off-the-shelf GPUs.
\item In this work we want to focus on scales of $\lesssim 50 {h^{-1}\rm\ Mpc})^3$.
On larger scales, perturbation theory and 2-point statistics of dark matter distribution
has been well studied.
\end{enumerate}

In the next two layers, we group these small-scale features together to extract the
large-scale features.
It is fair to say that, in the end, we mainly use the information of structures on
scales of $6-64\  h^{-1}\rm Mpc$.

The default architecture we describe in this section is closer to that used in
\cite{Mathuriya:2018luj} than the one used in \cite{Ravanbakhsh:2017bbi}.
In the next section we also discuss the effect of changes to this default architecture.

The structure of our neural network is shown in Figure \ref{fig_archi}.
It contains three convolution layers and three dense layers.
In the next subsection we discuss the implementation details.

\subsection{Convolution}
CNNs networks are designed to be ``shift/space invariance artificial neural
networks'', having shared-weights architecture and translation invariance
characteristics. They are especially suitable for analyzing images, videos, or any
kind of  structures with  a large number of pixels/voxels and shift/space invariant
properties.

The density field is fed to three convolutional layers.
The inputs of these layers are one or many cubes.
The convolutional kernels then convolve the inputs, and pass the results to the next layer.


The parameters in the convolutional kernels
decide features to be extracted from the input data.
They encode the prediction of cosmological parameters.
For example, the first layer contains 32 $3^2$-filters;
this means that 32 features are extracted,
by conducting dot product of the kernels of filters and the $(6\ h^{-1}\rm Mpc)^3$ sub-cubes
of the data, with a stride of $2\ h^{-1}\rm Mpc$.
Clearly, the information extracted here belongs to the highly non-linear clustering region.
The summation of the dot-products are transformed by the activation function (to have non-linear transformation in the network),
for which we use rectified linear unit (ReLU), $f(x)=\max(x,0)$.
This simple form enables fast calculation of gradients and effectively suppresses over-fitting,
and we accept it in the dense layers.

In Figures \ref{fig_layers1},\ref{fig_layers2}, we show how the CNN works.
Step by step, the features are extracted by the three layers, and become more and more condensed.
Different filters identify different features.
With a large number of filters we are able to perform a very comprehensive statistical analysis.
The final outputs are 128 $2^3$-voxel cubes.
The Figures clearly show that two different cosmologies lead to significantly different outputs.

Notice that these cubes do contain enough information for the measuring of $\sigma_8$.  
In the input of the CNN, all information is stored in the $(2\ h^{-1}\rm Mpc)^3$ voxles. 
The first convolution is conducted using 32 $3^3$ kernels, thus, 
in the feature maps generated by it, each voxel contains information in a volume of $ (6\ h^{-1}\rm Mpc) ^3$ 
(i.e., information within such a volume is mixed together after the first convolution). 
After a pooling and the second convolution, each volxel in the feature map then contains 
information spanning a volume of $ (14\ h^{-1}\rm Mpc) ^3$, 
whose scale is already larger than than 8 $h^{-1}\rm Mpc$. 
So, after another pooling, a third convultion, and a third pooling operation,
the final 128 $2^3$ cubes is definitely capable for the probing of $\sigma_8$ 
\footnote{This convince us that the final output of CNN has the ability of probing $\sigma_8$, 
but since the CNN is too complicated for us to understand,
we can only use tests justified (e.g., Figure \ref{fig_multi_cosmology} and \ref{single_cosmology}) 
to check whether the information of $\sigma_8$ is really stored in these cubes. }.

The parameters of filters are tuned in the training process
in a way that they can extract features which are closely related to the cosmological information.
The optimized CNN is far more complicated than
any traditional statistics (e.g., 2-point and 3-point statistics).
This enables more comprehensive data mining.


\subsection{Batch normalization and pooling}

A batch normalization layer is placed before each convolution layer.
Batch normalization is achieved through a normalization step
that fixes the means and variances of each layer's inputs.
It was initially proposed to solve ``internal covariate shift'' problem
\footnote{The distributions of the internal layers' inputs keep changing,
causing problems in the training process.},
and can also regularize the network such that it is easier to generalize.
It has become a widely-accepted technique
for improving the speed, performance and stability of the neural networks.

Results of each convolutional layer,
 are also passed to a ``pooling'' layer to decrease the sample size.
\cite{Ravanbakhsh:2017bbi} suggests using averaging pooling for LSS data,
so we adopt it as one of our default options of the network.
However, for our architecture we find that max-pooling shows better performance.

\subsection{Fully Connected Layers}


Outputs of the final pooling are flattened and passed to three fully connected
layers with 1024, 256 and 2 neurons, respectively. This system is a very complicated collection 
of non-linear mathematical functions, and is able to build up a connection between the features
extracted by the CNN (in our case the 128 $2^3$ cubes) and the values of $\Omega_m$ and $\sigma_8$.
To suppress over-fitting, we have a $20\%$ dropout layer placed before the dense layers.

We adopte MSE (mean squared error) as the loss function to describe the difference between the 
predictions of the whole neural network and the ``true'' values of ($\Omega_m$, $\sigma_8$).
By default, we use the Adaptive Moment Estimation (Adam)  optimization algorithm
\citep{2014arXiv1412.6980K} to find the values of parameters (of the CNN and the fully connected layers)
which minimize the loss function.




\section{Results}\label{sec:results}

In this section we present the results of the neural network.

\subsection{Convergence test}

The leftmost column of Figure \ref{fig_lc}
shows the {\it learning curves} of two different runs using the default architecture.
Plotted are the average of the predictions from the  500 single-cosmology samples.
The two runs yield very different predictions at the early stage of training,
while after $\sim$200 epochs they start to converge and
yield similar predictions ($n$ training epoch means the whole training samples 
are fed to the network by $n$-th time).
After 400 epochs, their predictions are basically the same.

Thanks to the lightness of our neural network 
the training is not significantly computationally expensive.
It can finish within 1 week using the CPU of a personal computer.
This makes it useful for many cosmologists 
who are interested in machine learning
but not familiar with multiple-GPU implementations.

\subsection{Different architecture}

The middle columns of Figure \ref{fig_lc} show some tests
on the architecture choices.

We tuned the architecture by decreasing, either the number of filters in the convolutional layers,
or the neurons in the dense layers, by a fraction of 50\%;
yet we find no significant change in their learning curves
We also tried doubling these numbers, and still obtain similar learning curves.

In the middle-right panel, we present results when we 
1) use max-pooling instead of average-pooling;
2) use stochastic gradient (sgd) as the optimizer (the default optimizer is Adam).
These changes slightly improve the performance (especially, decreasing the bias in
the estimation of $\Omega_m$).

\subsection{Bias correction}

We find a bias in the estimated parameters.
This bias is smaller than the one reported by \cite{Ravanbakhsh:2017bbi}, but larger than the apparently unbiased results of \cite{Mathuriya:2018luj}.
In most cases, we underestimate $\Omega_m$ by about 0.005 (less than $2\%$).
While $\sigma_8$ is under-estimate by about 0.02 which is about $2.5\%$.
Increasing the training epochs to 1,500 does not reduce this bias.

We do not have a definitive answer for the origin of this bias.
One possibility is that it comes from the limited power of the dense layers
in regressing the cosmological parameters from the 128 $2^3$-voxel features.
This hypothesis is supported by the fact that placing another 512-neuron layer after the
256-neuron layer, to improve the ability in mapping the many voxels to the
parameters, the bias is obviously decreased (see rightmost panel of Figure
\ref{fig_lc}). 

To have a better understanding of the bias,
we plot its parameter-dependence in Figure \ref{fig_multi_cosmology}.
We find a clear trend of increasing bias
at larger values $\Omega_m$ or $\sigma_8$.
This trend is again consistent with the results in \cite{Ravanbakhsh:2017bbi}.

Adding more layers/neurons in the dense layers
to further decrease the bias goes against of our objective
of having a simple and light convolutional network.
Instead, we opt for
a simpler (and possibly more accurate) treatment by
deducting the bias based on a polynomial regression
\footnote{A polynomial regression may sound arbitrary,
but in principle it has no intrinsic difference from a mapping using dense layers.}.
Figure \ref{fig_bias} shows that the biases
can be well estimated using a 3-rd order polynomial
as functions of $\Omega_m$ and $\sigma_8$.

The fitting formula (here a high order polynomial) may become complicated 
when there are 6-7 model parameters, 
however implementing it is always simple
and straightforward. Also, its complexity is not comparable with that of the
neural network. 


\subsection{Cosmological constraint}

Figure \ref{single_cosmology}
shows the final constraints derived from the 500 single-cosmology samples,
where the bias has been subtracted based on the polynomial regression.
To avoid self-correction, the regression is derived using the multi-cosmology samples,
which have no overlapping from the single-cosmology samples.

We find the CNN accurately predicts the parameters as
\begin{equation}
\Omega_m=0.3073\pm0.0015,\ \sigma_8=0.8178\pm0.0029.
\end{equation}
They are statistically consistent with the ground truth
$(0.3071,\ 0.8228)$.
We find the prediction of $\sigma_8$ still suffers from a $\approx$1$\sigma$ bias;
this can be overcome by performing a more precise
bias-estimation based on larger amount of samples
(e.g. a point-by-point correction on the grid)
\footnote{The bias on $\sigma_8$, being on level of 1$\sigma$, 
is not statistically significant and worth further studies. 
One can use multiple such realizations, or a larger realization, 
to achieve better estimations.}.

The statistical error of $\Omega_m$ is 6 times smaller than
the Planck 2015 TT,TE,EE+lowP+lensing constraint,
4 times smaller than Planck+BAO+JLA+$H_0$ constraint
\citep{ade2016planck} 
\footnote{The comparison is not ``very suitable'' since 1) Our analysis
is based on noise-free simulations; 2) We do not include the systematical 
uncertainties; 3) CMB and LSS are very different types of observations.
This comparison is just for illustrative purpose to 
enable readers easily understand that the results are really precise 
and the CNN analysis of LSS data is promising.}.
Having derived this result from a $($256 $h^{-1}\ \rm Mpc)^3$, 2 $h^{-1}\ \rm Mpc$ 
resolution sample shows the great potential of using neural network to estimate cosmological parameters from the LSS.

One caveat is that the variance of parameters 
may depend on the value of parameters. 
In practice, one can generate several sets of mocks on 
different positions of the parameter space to estimate this effect. 
Then the dependence on the whole parameter space 
can be modeled via interpolation.


Compared with the results of \cite{Ravanbakhsh:2017bbi},
the errors of our predicted $\Omega_m$ and $\sigma_8$ are 
5 and 2 times smaller,
while our constraints are achieved using simulation samples 
with 8 times smaller box-size and 
64 times smaller number-of-particles
\footnote{If we simply assume that the information scales with the number of particles,
then our CNN is 40/16 times better in predicting $\Omega_m$/$\sigma_8$ compared with \cite{Ravanbakhsh:2017bbi}.}.

As a comparison with traditional methods, 
\cite{Ravanbakhsh:2017bbi} conduced a power spectrum analysis,
and found the error of its predicted $\Omega_m$ is 
2.6 times larger than their CNN error,
hence 13 times larger than our CNN error.
If we assume the accuracy of power spectrum analysis 
scales with the square-root of sample volume,
then our CNN is $\approx25$ times more precise 
than a power spectrum analysis in predicting $\Omega_m$.

To better understand the potential of the CNN 
we also made a comparison with the 
2-point correlation function (2pcf) analysis results.
The 2pcf constraints on parameters are derived 
by measuring the shape and amplitude
of the 2pcfs using samples in the many cosmologies,
to build an emulator. 
We find CNN constraints on $\Omega_m$/$\sigma_8$ are 
3.5/2.3 and 19/11 times more precise than 
the 2pcf analysis using the clustering range of 
0-130 and 10-130 $h^{-1}$ Mpc, respectively.
Notice that the 2pcf analysis is very ideal, 
since in realistic analysis we usually use 
the clustering region s$\gtrsim30 h^{-1}$ Mpc,
while the amplitude information, being affected by many systematics,
can not be easily utilized.

\begin{table*}
\footnotesize{
\caption{\textrm{Comparison between this work and \cite{Ravanbakhsh:2017bbi} }}
\begin{center}
\label{tab1}
\begin{tabular}{|c|c|c|c|}
  \hline
  & & & \\
   & Method & $\rm Training\ sample$ & $\rm Relative\ error$ of $(\Omega_m, \sigma_8)$\footnote{Defined as $\Delta y/y$ ($y$ stands for $\Omega_m$, $\sigma_8$)
  where $\Delta y$ includes both the statistical error and bias}  \\
  & & & \\
  \hline
   & & 450 simulations &    \\
  Ravanbakhsh et al. \citep{Ravanbakhsh:2017bbi}  & CNN & $(512\ h^{-1} \rm Mpc)^3$, $512^3$ particles  & (0.028, 0.012) \\
  \hline
  & & 450 simulations &    \\
  Ravanbakhsh et al. \citep{Ravanbakhsh:2017bbi}  & {Power spectrum} & $(512\ h^{-1} \rm Mpc)^3$, $512^3$ particles  & (0.072, 0.013) \\
  \hline
  & & 465 simulations &    \\
  This work  & CNN & $(256 h^{-1} \rm Mpc)^3$, $128^3$ particles  & (0.0048, 0.0053) \\
  \hline
  & & 465 simulations &     \\
  This work  & 2pcf, $s \in (0,130) h^{-1} \rm Mpc$ & $(256 h^{-1} \rm Mpc)^3$, $128^3$ particles  & (0.017, 0.012) \\
  \hline
  & & 465 simulations &  \\
  This work  &  2pcf,  $s \in (10,130) h^{-1} \rm Mpc$ & $(256 h^{-1} \rm Mpc)^3$, $128^3$ particles  & (0.1, 0.06) \\
  \hline
\end{tabular}
\end{center}
  }
\end{table*}

\subsection{Error tolerance}

\begin{figure*}
   \centering
    \includegraphics[width=16cm,height=22cm]{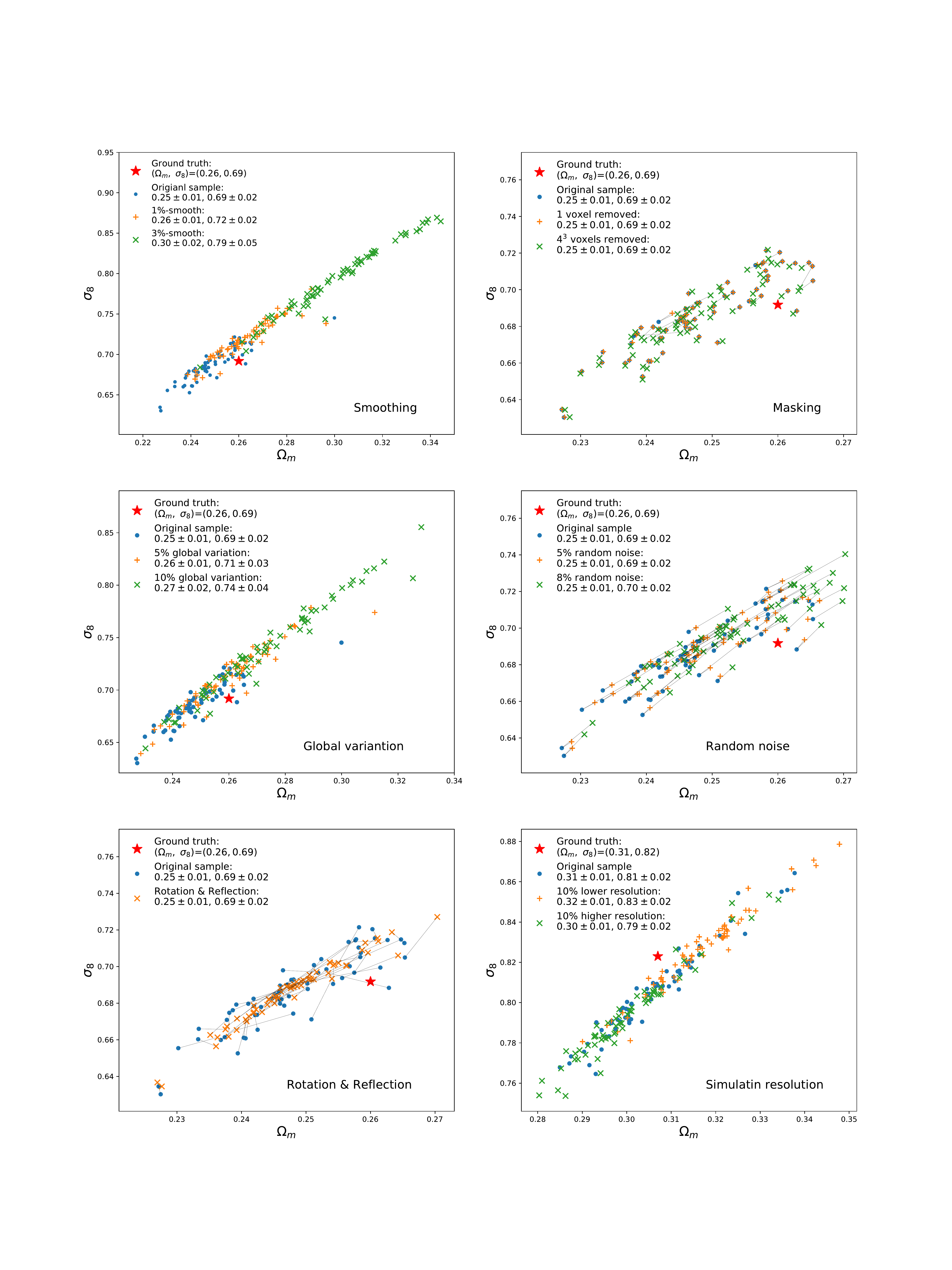}
   \caption{\label{fig_ET}
   Error-tolerance tests.
   A 3\% smoothing or 10\% global variation leads to considerable change in the predicted results
   ($\sim 2 \sigma $ shift in central values, $~\sim 100\%$ enlarged errors).
   1\% smoothing, 5\% global variation, and 10\% change in the simulation's resolution
    mildly affect the prediction ($\sim 1\sigma$ shift in central values, errors unchanged).
   Other cases, including the 1 or $4^3$ voxels removal,
   5\% or 8\% random noise addition,
   rotation and relfection,
    does not affect the results at all.
   }
\end{figure*}

So far we only apply the neural network to \emph{ideal} datesets --
density fields regularly sampled in a 3-d grid based on the dark matter particles.
In reality, the data obtained in observations contain many sources of systematics.
Here, we quantify how this noise affects the performance of the neural network. 

The ET (error tolerance) tests are presented in Figure \ref{fig_ET}.
For simplicity, in these tests we only use {\it one} $128^3$-voxel sample,
generated using $(\Omega_m,\ \sigma_8)$=(0.26, 0.69).
We split the grid into 64 $32^3$-voxel subgrids to obtain 64 sets of estimated parameters.
Adding different kinds of noise into the subgrids,
and feed them to the neural network to predict parameters.
When a certain kind of noise was added,
we check whether the estimations are changed,
and get some understandings about the effect of noise.

In summary, we find that:
\begin{itemize}
 \item A smoothing of the sample
 \footnote{Our smoothing means that each voxel is replaced by a weighted sum of itself and its six nearest neighbors.
 Different types of smoothing can have different effect and should be tested individually.}
 can lead to disastrous effect.
 Even a 1\% smoothing shifts the estimation by $\approx$2$\sigma$.
 A 3\% smoothing doubles the shifts and also doubles the statistical scattering
 \item In contrast, the performance of the neural network is very robust to missing voxels.
 We mask 1 or $4^3$ voxels in each of the $32^3$-subgrid
 (by setting their values to 0), and find the predicted results almost unchanged.
 This ET ability is helpful, since in real observations
 there are always many masked regions.
 \item The performance is not significantly improved
 if we conduct data enhancement (DE)
 via rotation and reflection.
 The number of the 3d subgrids can be increased by as much as 48 times
 after DE.
 No significant improvement in the predictions is detected
 if we feed the 48-times more samples to the neural network.
 \item The predictions are very robust to Gaussian noise.
 In this test, all voxels are multiplied by a Gaussian random variable with
   a standard deviation of 5\% or 10\%.
 The central values and errors remain unchanged.
 \item  If we introduce a 5\% or 10\% {\it global} variation (rescaling) 
 of the density field
 (linearly increased from 0\% at $x=0$ to the maximal
 value at $x=256 {h^{-1}\rm  Mpc}$),
 notable change appears in the predicted results.
 Thus, when analyzing observational data, one should be careful about the factors
 which can globally change the survey properties in a large area.
 \item In case that we feed the neural network using samples produced in 10\% lower/higher resolution
 (decreasing/increasing the number of simulation particles by 10\%),
  the central values are mildly shifted ($\sim 1\sigma$).
\end{itemize}

The above tests have not previously been performed in 
\cite{Ravanbakhsh:2017bbi,Mathuriya:2018luj},
so this constitutes a  first 
check of these systematics. 
The results justify that the neural network analysis is not, 
as some suspected, 
significantly sensitive to even tiny systematics variations.
Although these tests are oversimplified compared with those that should be done 
when dealing with real observational data,
the tests enable us to have some preliminary understanding 
of the influence of the systematics effects.

When handling real observational data, an observational artifact can be overcome in two ways.
1) Designing a neural network that is robust to it.
This can be done by modeling the systematics by several parameters,
and allowing them to run over a wide range in the training sample,
so that the neural network is adaptable to a wide range of systematics parameters.
This can handle those systematics which are not well understood.
2) For well-known systematics, 
one can simply add it into the training sample, 
so that its effect is considered by the neural network 
in the training process.

\section{Concluding Remarks}\label{sec:conclusion}

We used a deep convolutional neural network to estimate cosmological parameters from simulated dark matter distributions.
The simulations are $128^3$-voxel, $(256\ h^{-1}\ \rm Mpc)^3$ cubes of the dark matter density contrast field.
The neural network, designed to have three convolution layers, three dense layers, including batch normalization and pooling layers,
builds up a connection from the field to the cosmological parameters.
It is able to yield accurate prediction of the cosmological parameters
after $\sim200-300$ epochs of training.
We also studied some variations on the architecture to test its convergence and overall performance.

In the estimated parameters, we find a persistent bias that can not be resolved by increasing the training epochs.
We believe that this bias arises from the limited power of the dense layers,
which are responsible for mapping the outputs of the convolution to the cosmological parameters.
Using more sophisticated dense layers,
or simply applying a subtraction based on polynomial regression,
the bias can be suppressed.
We also tested the error-tolerance abilities of the neural network,
including the abilities against smoothing, masking, random noise, global variation, rotation, reflection and resolution.

The robustness tests are still preliminary and only enable us 
have some basic understanding about the influence of the systematics.
Once one uses dark matter distributions to populate galaxies, the inclusion of
more complicated systematics would be required due to the complexity of the problem. 
This needs to be explored in future analysis.
Also, considering that the size of the sample 
used in the test is relatively small,
we can only obtain some basic understanding of the systematics at this point.

We obtain precise estimations, with statistical scattering of $\delta \Omega_m$=0.0015 and $\delta \sigma_8$=0.0029, from the neural network.
The statistical error of $\Omega_m$ is 6 and 4 times smaller than
the Planck and Planck+ext constraints presented in \cite{ade2016planck}.
We conclude that deep neural networks are very promising in estimating cosmological parameters from the LSS.

The persistent bias in the prediction of our neural network would be the biggest caveat
limiting the power of the technique.
The bias was also detected in the work of \cite{Ravanbakhsh:2017bbi},
yet the authors did not provide a strategy to overcome it.
It seems that the bias is greatly reduced if one uses a more complex network architecture with seven convolutional layers and $128^3$ voxels as an input \citep{Mathuriya:2018luj}.

The approach that we develop to correct the bias is a simple subtraction based on polynomial regression.
This is not completely satisfactory and future work should aim to address this problem, i.e. measuring how it depends on the architecture parameters.
This will allows us to design better architectures with a smaller bias,
and conducting more concrete tests based on larger training samples.
This study is a required prerequisite to conduct a reliable, comprehensive analysis of LSS using deep learning.

On the physical side there are at least many directions for future work.

\begin{enumerate}
\item In this simple work, we haven't consider the role of redshift space distortion
(RSD) in the parameter estimation.
We tend to believe that the RSDs, which creates more cosmological dependent
features in the matter distribution,
should lead to better parameter estimation.
We will test this supposition in forth-coming works.

\item In the case of a survey covering a $(512\ h^{-1}\ \rm Mpc)^3$ or $(1\ h^{-1}\ \rm Gpc)^3$
volume of density field, one can further decrease the statistical error by 3 or 8 times.
In that case the bias and error tolerance (to systematics)
of the neural network would be essentially important.
Lightcone effect, selection function, galaxy bias, redshift errors, or even barynonic effects,
should be tested in certain circumstances.

\item The resolution of our input sample, $2\ h^{-1}\ \rm Mpc$, is a bit high when
considering the
current and near-future spectroscopic surveys, which have low comoving number
densities.
So it will be necessary to apply the method to lower-resolution,
more realistic galaxy samples.
In the next step, we will apply it to dark matter halo samples,
and see whether the neural network is still able
to achieve precise parameter estimation in such circumstances.

\item While in this work we only consider
the predictions of $\Omega_m$ and $\sigma_8$,
in general the CNN can be used to probe
any cosmology or astrophysical parameters
which can affect the large scale structure.
An incomplete list include the parameters related with the Hubble constant
\citep{Riess2016,Wang2017},
dark energy equation of state and its time dependence
\citep{miao2011dark,ZW2018,LFZZ2019},
gravity \citep{Sotiriou2010,Tsujikawa2010,Abbott2017,ZWW2018,LCH2019,Zhang22019},
galaxy formation and evolution  \citep{White1977,Dressler1980},
and so on.

\end{enumerate}

\acknowledgments

We thank Kwan-Chuen Chan, Cheng Cheng, Zhiqi Huang, Yin Li, Guangcong Wang
and Xin Wang for helpful discussions.
XDL thanks Ziyong Wu and Xiaolin Luo for kind helps.
We acknowledge the use of {\it Kunlun} cluster located in School of Physics and Astronomy, Sun Yat-Sen University.

XDL acknowledges the supported from NSFC grant (No. 11803094),
the Science and Technology Program of Guangzhou, China (No. 202002030360).
J.E. F-R acknowledges support from COLCIENCIAS Contract No. 287-2016,
Project 1204-712-50459.  
CGS acknowledges financial support from the National Research Foundation
(NRF; \#2017R1D1A1B03034900, \#2017R1A2B2004644 and \#2017R1A4A1015178).
ZL was supported by the Project for New faculty of Shanghai JiaoTong University
(AF0720053),
the National Science Foundation of China (No.  11533006, 11433001)
and the National Basic Research Program of China (973 Program 2015CB857000).


\bibliographystyle{aasjournal}

\end{document}